\documentclass[journal, 10pt, twocolumn]{IEEEtran}

\usepackage{cite} 
\usepackage{tikz}
\usepackage{amsmath,amsfonts,amssymb,amsthm}

\usepackage[utf8]{inputenc}
\usepackage[nottoc]{tocbibind}
\usepackage{hyperref}
\usepackage{caption}
\usepackage{subcaption}
\usepackage{dsfont}
\usepackage{xspace}
\usepackage{multirow}
\usepackage{gensymb}
\usepackage{cuted}

\usepackage{enumitem}
\usepackage{algorithm}
\usepackage{algorithmicx}
\usepackage[noend]{algpseudocode}

\newcommand{\lu}{\lambda_U}

\newcommand{\one}{\mathds{1}}

\newcommand{\B}{\mathcal{B}}
\newcommand{\D}{\mathcal{D}}
\newcommand{\U}{\mathcal{U}}

\newcommand{\newInfo}[1]{\textcolor{black}{#1}}   

\newcommand{\lotte}[1]{\textcolor{black}{#1}}

\begin{document}
\title{\textsc{beam-align}: distributed user association for  mmWave networks with multi-connectivity}

\author{Lotte~Weedage, 
        Clara~Stegehuis, and 
        Suzan~Bayhan 
\thanks{Authors are with the Faculty
of Electrical Engineering, Mathematics and Computer Science~(EEMCS), University of Twente, The Netherlands.\protect\\
Corresponding author's e-mail: l.weedage@utwente.nl.}
}
\maketitle
\begin{abstract}
Since the spectrum below 6 GHz bands is insufficient to meet the high bandwidth requirements of 5G use cases, 5G networks expand their operation to mmWave bands.
%
However, operation at these bands has to cope with a high penetration loss and susceptibility to blocking objects. 
%
%
%
Beamforming and multi-connectivity~(MC)  can together mitigate these challenges. But, to design such an optimal user association scheme leveraging these two features is non-trivial \newInfo{and computationally expensive}. 
Previous studies either considered a fixed MC degree for all users or overlooked beamforming. Driven by the question \textit{what is the optimal degree of MC for each user in a mmWave network,} we formulate a user association scheme that maximizes throughput considering beam formation and MC.  
Our numerical analysis shows that there is no one-size-fits-all degree of optimal MC; it depends on the number of users, their rate requirements, locations, and the maximum number of active beams at a BS.
%
Based on the optimal association, we design \textsc{beam-align}: an efficient heuristic with polynomial-time complexity $O(|\mathcal{U}|\log|\mathcal{U}|)$, where $|\mathcal{U}|$ is the number of users. Moreover, \textsc{beam-align} only uses local BS information \newInfo{ - i.e. the received signal quality at the user}. \newInfo{Differing from prior works, \textsc{beam-align} considers beamforming, multiconnectivity and line-of-sight probability.}
%
Via simulations, we show that \textsc{beam-align}
performs close to optimal in terms of per-user capacity and satisfaction  while it outperforms frequently-used signal-to-interference-and-noise-ratio based association schemes.   
We then show that \textsc{beam-align} has a robust performance under various challenging scenarios: the presence of blockers, rain, and clustered users. 
\end{abstract}




\section{Introduction}
\begin{figure}
    \centering
    \includegraphics[width = 0.4\textwidth]{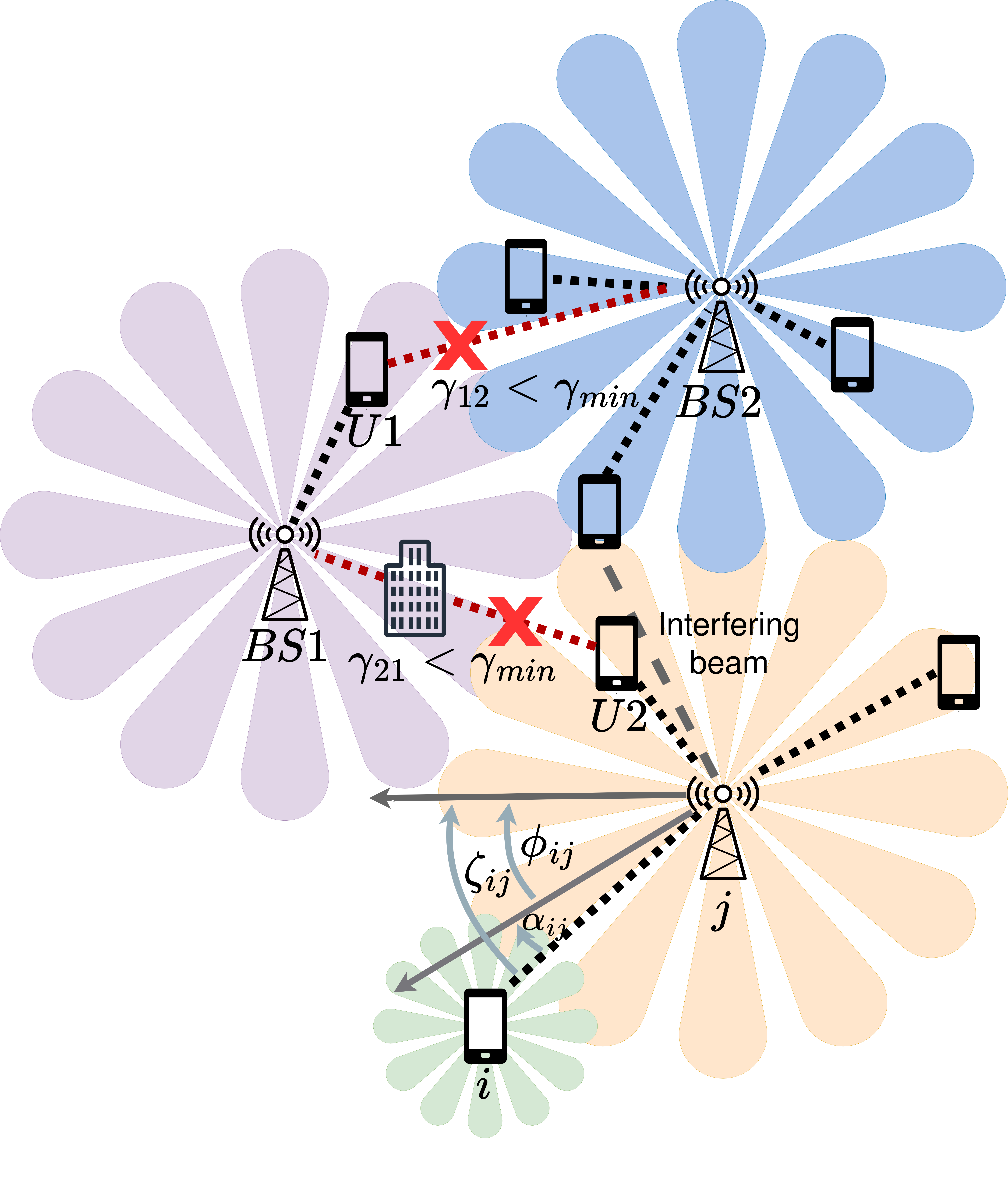}
    \caption{A mmWave scenario in which 
     users and BSs have directional antennas and 
    users can connect to multiple BSs. Links might be blocked by e.g. a building or another user. The beamforming gain depends on the boresight angle $\phi_{ij}$, geographical angle $\zeta_{ij}$ and misalignment angle $\alpha_{ij}$ of user $i$ and BS $j$. A user can only have a connection with a BS if the link-SNR 
    is higher than $\gamma_{\text{min}}$. 
    }
    
    \label{fig:mmwave_scenario}
\end{figure}

5G networks can use frequency bands above 6 GHz known as mmWave bands to meet the high rate requirements of data-intensive use cases such as immersive media applications~\cite{sakaguchi2017and}. However, these high frequency bands have increased path losses compared to sub-6GHz channels and are more susceptible to channel impairments, e.g. due to blocking objects or raindrops~\cite{liu2016user}. 
Massive directional antennas in mmWave base stations (BSs) can compensate for this increased path loss, by transmitting in focused beams via \emph{beamforming}. Beamforming increases the channel gain based on the alignment of the beams of a BS and a user \cite{nadeem2010analysis}. Alignment then becomes a key factor in determining the link quality.
Additionally, to compensate for blockage or other channel impairments such as rain, users can sustain a higher connection reliability by simultaneously connecting to multiple BSs as illustrated in  Fig.\ref{fig:mmwave_scenario}.
While \emph{multi-connectivity} (MC) can be implemented in different ways, e.g. packet duplication or load balancing~\cite{Suer2020Multi-ConnectivityOverview},
we investigate a scenario with \textit{load balancing} MC, in which a user's downlink traffic is split over different links to also increase the user throughput.

With beamforming and MC, 5G networks can unlock the potential of mmWave bands and serve their users with high data rates. However, two questions arise in this context: (i) what is the optimal degree of MC for a user to attain the minimum data rate asserted by its application? and (ii) to which BSs should each user connect, i.e. how to design the user association (UA) scheme?
These questions have also been raised by some prior studies, such as \cite{Gapeyenko2019OnDeployments, zhang2020optimal, weedage2021impact, petrov2017dynamic}. However, prior research on MC investigates the increased throughput or reliability offered to a single user~\cite{Gapeyenko2019OnDeployments} and mostly ignores the peculiarities of mmWave networks such as directional operation \cite{petrov2017dynamic}. 
%
Our earlier study in \cite{weedage2021impact} investigates a fixed MC scheme in which each user is associated with $k$ BSs and 
concludes that the number of connections per user should be dynamically determined. Under a $k$-MC scheme, per-user capacity decreases while the network can attain higher reliability in terms of a reduced outage probability in comparison to single connectivity~(SC) settings.
However, finding the optimal degree of MC for each user requires solving an optimization problem considering all users' rate requirements and channel states. Acquiring this information is challenging and impractical, making optimal schemes prohibitively complex.  

Based on the aforementioned challenges, our goal is to design a computationally-efficient UA scheme that maximizes network throughput while meeting the users' minimum rate requirements by exploiting a dynamic form of MC. To the best of our knowledge, our study is the first dynamic MC scheme that takes advantage of both beamforming and MC \newInfo{and only uses local information for the user's decision to connect to a BS or not}.
In this paper, we address the following research questions:
\begin{itemize}
    \item What is the optimal number of connections in a mmWave network for a user with a certain downlink rate requirement? How does the optimal number of connections depend on the network load? 
    \item How does the transmission beamwidth of a BS (e.g., smaller or larger) affect the optimal number of connections per user and the network performance? 
    \item Can we design a near-optimal distributed UA scheme that is only based on local user information?
    \item How do realistic scenarios such as blockage, rain and user clustering affect the performance of UA schemes in terms of per-user throughput and user satisfaction level?
\end{itemize}

While addressing these questions, we have the following key contributions in this paper:
\begin{itemize}
    \item We propose an optimal UA scheme based on maximizing a utility function of the throughput. In contrast to previous research such as \cite{xue2023user} and \cite{cai2021joint}, this UA scheme takes the key properties of mmWave bands into account, such as beamforming \newInfo{with beam scheduling} and increased path loss \newInfo{with blockage}, \newInfo{and is based on dynamic MC instead of fixed MC or SC}.
 We show that in the optimal UA scheme, most users have MC when the network load is low. Moreover, our analysis of this optimal UA scheme shows that the decision to connect to a certain BS is primarily based on the antenna alignment between the BS and the user. Furthermore, we conclude that there is no one-size-fits-all degree of MC; the optimal degree depends on the network load, user rate requirements, number of active beams, and antenna orientation.

    \item Different from studies such as \cite{Gapeyenko2019OnDeployments} which assume that users associate to the BSs offering the highest SNR, we devise \textsc{beam-align} based on our observation that optimal UA results in users connecting to the BSs that have a small misalignment with the user's receive antenna. In \textsc{beam-align}, users are associated to BSs based on the alignment of the beams. This heuristic offers benefits in terms of per-user throughput and satisfaction levels, which are close to optimal, e.g., an average per-user capacity that is between $2.0\%$ and $19.1\%$ lower than the optimal scheme for low user densities. Still, it only has polynomial time complexity, $O(|\mathcal{U}|\log|\mathcal{U}|)$ where $|\mathcal{U}|$ is the number of users in the network and only requires knowledge of the antenna orientation of a user. Our numerical analysis validates that only considering SINR leads to a significant loss in capacity.
    
    \item We analyse the robustness of \textsc{beam-align} as well as SINR-based association schemes in the literature under various scenarios: namely, when i) there are blockers; ii) there are different levels of rain that impair the mmWave links; and iii) the users are clustered. Moreover, we compare our heuristic to the state of the art (MOHHO, \cite{jin2022toward}) and show that \textsc{beam-align} outperforms MOHHO both in terms of per-user capacity and energy efficiency, \newInfo{while \textsc{beam-align} does not even optimize for energy efficiency}. In comparison to MOHHO and other heuristic user association schemes, \textsc{beam-align} is more computationally efficient and can be run on scenarios with large numbers of users.
\end{itemize}

In the following sections, we first overview the most relevant work in the areas of MC, mmWave and user association (Section~\ref{sec:literature}). Next, we elaborate on the scenario and models we use in this paper (Section~\ref{sec:system_model}) and propose an optimization model for UA, based on maximizing the total throughput (Section~\ref{sec:optimization}). Then, we analyze the results of the optimal UA scheme and present \textsc{beam-align} (Section~\ref{sec:simulations}), which we compare to the optimal scheme and other heuristic UA algorithms. In Section~\ref{sec:resilience}, we investigate the robustness of both UA schemes. Finally, we discuss the limitations of our work and some future directions in Sections~\ref{sec:discussion} and~\ref{sec:conclusion}.

\section{Related work}\label{sec:literature}
In this section, we describe related work on UA in a mmWave setting. We focus on four areas: SC user association, MC user association, beamforming and simple heuristics for UA. For a complete overview of 5G user association, we refer to \cite{liu2016user} and \cite{attiah2020survey}.

\textit{SC user association:} Optimal UA under SC for mmWave networks is widely studied in the literature~\cite{cacciapuoti2017mobility}. 
Existing studies design new algorithms for UA, for example by using a Markov decision process \cite{mezzavilla2016mdp} to minimize the number of handover decisions and to maximize throughput. However, a key shortcoming of these works is that they neglect the directionality gain of beams, while this simplification might significantly influence the performance of a UA scheme. 
In \cite{kim2018millimeter}, users decide to connect to a BS based on the directional signals in an area with a high building density, as buildings can reflect signals in different directions. 
In this line of research, the optimization criterion is typically the total network throughput. Other objectives that have been investigated are the energy efficiency~\cite{paul2020boost} or outage probability~\cite{zhang2020optimal}. Another study worth mentioning is \cite{shokri2015user}, in which the authors formulate and solve an optimization problem based on maximizing throughput for different beamwidths. All UA schemes that are described above can be extended to a fixed MC UA scheme, where all users connect to $k$ BSs. However, extending them to a dynamic MC scheme, which decides the optimal level of MC for each user, is not straightforward or might incur high complexity.

\textit{MC user association:}
Earlier studies on MC can be divided into two categories: \textit{fixed} MC \cite{liu2020user, liu2021user, liu2020resource, perdomo2020user}, where the number of connections per user is fixed, and \textit{dynamic} MC, where each user can have a different number of links \cite{ozkocc2021impact, sopin2022user}. Generally speaking, the main objective in these studies is to maximize total throughput \cite{liu2021user, liu2020resource, ozkocc2021impact}, energy efficiency \cite{perdomo2020user, chen2022joint, gao2023wip, wang2023user}, minimizing the number of resources in the system~\cite{tatino2020learning}, or minimizing outage probability \cite{sopin2022user}. 
The closest studies to ours are \cite{tatino2018maximum, Gapeyenko2019OnDeployments, petrov2017dynamic, wang2023user}. Tatino \textit{et al.}~\cite{tatino2018maximum} consider a scenario with MC and compute a near-optimal UA scheme based on column generation, showing that MC can drastically increase throughput and reliability. In contrast to \cite{tatino2018maximum}, we take beamforming into account, as this is a key property of mmWave. Petrov \textit{et al.} \cite{petrov2017dynamic} provide a framework based on queuing theory for mmWave scenarios, including MC and focus on small scenarios with a high building density. Their conclusions are also similar; MC increases the throughput and reliability. Our study differs from \cite{petrov2017dynamic} in the fact that in \cite{petrov2017dynamic}, beamforming is not explicitly used. Moreover, the notion of MC is different: in \cite{petrov2017dynamic} under MC, a user can reconnect to a different BS after blockage while we assume that users have multiple connections at the same time under MC. \newInfo{In \cite{wang2023user}, the authors heuristically solve the throughput maximization problem by decoupling the power allocation and user association problems. However, their proposed UA algorithm is still computationally expensive and can only be solved for small instances}. Research questions raised in \cite{Gapeyenko2019OnDeployments} are similar to our questions: \textit{what is the best degree of MC?} While motivated by the same research question, our approaches and focus differ. The research in \cite{Gapeyenko2019OnDeployments} overlooks the directional transmissions and assumes a fixed UA scheme in which each user selects a fixed number of BSs based on the highest SNR. On the contrary, we aim at computing the \textit{optimal degree of MC for each user jointly} based on their rate requirements, position in relation to the other users and the BSs, as well as beams of the neighboring BSs.  Moreover, our study suggests that a highest-SINR-based MC might not yield as high performance as our proposal would offer. Additionally, we conclude that there is no one-size-fits-all degree of MC; the optimal degree depends on the network load, user rate requirements, beamwidth of operation, and antenna orientation.
Sopin \textit{et al.}~\cite{sopin2022user} investigate different UA schemes in joint terahertz and mmWave networks with beamforming and MC. However, this study focuses on minimizing the outage probability due to blockages and mobility of users, while our focus is on optimizing throughput and satisfaction of users.

\textit{Beamforming:}
Directional beams in mmWave channels help overcome low signal quality due to high path loss. Several studies investigate how to steer, manage and align these beams to achieve the highest throughput or best efficiency \cite{aghashahi2021stochastic, kanhere2021performance, tsai2021high}. All of these works show that operation with smaller beams leads to higher throughput. However, smaller and more beams may also result in higher overhead and complexity, as all these beams have to be managed. Furthermore, smaller beams may not perform well when there is some uncertainty in the users' positions. As directional beams cause different gain patterns around a BS, UA under beamforming becomes a difficult problem that can be solved in different ways. In \cite{alizadeh2020distributed}, the mmWave UA problem is formulated as a matching and auction game, respectively, and in \cite{liu2020user, tatino2020learning, gao2023wip, chongrui2023deep} the authors use machine learning to solve the UA problem. Due to the complexity of the problem, most of these models focus on a small setting with only a few users, while mmWave is used for scenarios with high user density. Moreover, most of these works are based on SC, as MC brings even more complexity and overhead. Our work differs from earlier works as we consider a more complete mmWave setting with both beamforming and MC. 

\textit{UA heuristics:}
Several learning-based UA heuristics are proposed in \cite{tatino2020learning} which do not assume perfect channel knowledge. The authors compare their proposals to two baseline (distributed) heuristics: connecting to all available BSs and connecting to as many BSs as necessary to satisfy the users' quality-of-service constraints. 
Another study is \cite{gupta2021stability} which compares four simple distributed UA policies in a setting where users join and leave the network. They show that UA should be based on the highest throughput. However, MC and beamforming are not considered.
Three heuristics for UA are studied in \cite{mezzavilla2016mdp}: (1) connect to the least-loaded BS, (2) connect to the highest instantaneous rate, and (3) connect to the highest SNR. In this work, the authors showed that out of these three heuristics, the SNR-based approach performed best in terms of spectral efficiency. However, some of these UA schemes require knowledge of the entire network. Contrary to most of these works that only verify their results on small instances, we aim to design a UA scheme that is scalable and performs well also in larger settings. Lastly, a framework for UA is proposed in \cite{paul2020boost} based on minimizing the transmission time and the power budget of a BS, by using a clustering algorithm to cluster users in the same beam. However, this research does not take into account the possibility for users to have MC. \newInfo{Several heuristics considering MC are proposed that optimize energy efficiency \cite{cai2021joint, jin2022toward} or throughput \cite{chen2023resource}}. However, unlike our study, these devised heuristics overlook beam directionality.

In a nutshell, our work differs from the previous studies in that we investigate the optimal degree of \textit{dynamic} MC for mmWave networks with directional transmission, where users might have a different number of connections based on their data rate requirements.
We provide an optimal UA scheme that is used to design a distributed heuristic that performs near-optimally with only $O(|\mathcal{U}|\log|\mathcal{U}|)$ complexity where $|\mathcal{U}|$ is the number of users in the network.

\section{System model}\label{sec:system_model}
We consider a cellular network consisting of a set of BSs $\B$ operating at mmWave bands which are expected to serve a set of users $\U$. BSs are deployed on a hexagonal lattice.
If there is a link between user $i \in \U$ and BS $j \in \B$, we refer to this link as link $ij$.
Each user has a downlink rate requirement $R_{\text{min}}$ to have an acceptable service quality. Given that mmWave bands have abundant spectrum, we assume that neighboring BS operates over a different channel with a bandwidth of $W$ Hz. We use a frequency re-use scheme with re-use factor $7$, which is a common re-use factor in hexagonal BS deployment.

In the following, we introduce the considered system model and key assumptions of our work.

\subsection{Antenna model}
To calculate the antenna gain for a transmitter or receiver operating with beams, we use the model proposed in mmWave-based IEEE 802.15.3c standard~\cite{toyoda2011antenna}.
Let us denote by $\theta^u$ and $\theta^b$ the beamwidth of a user and a BS, respectively. Moreover, $\phi_{ij}$ is the \textit{boresight} and $\zeta_{ij}$ the \textit{geographical} angle of link $ij$ as illustrated in Fig.~\ref{fig:mmwave_scenario}.
We assume that BSs and users have a sectorized linear antenna \cite{7880676, 7528494}, which means that the beams are placed uniformly around the BS and user. This results in the boresight angles
$\phi_{ij} = d \cdot \frac{2\pi}{\theta} $, where $d \in \D = \{1, 2, \ldots, m\}$ is the set of all beam directions. We refer to $\D^u$ and $\D^b$ as the set of user and BS beam directions, with beamwidths $\theta^u$ and $\theta^b$, respectively. 
{We assume that the number of RF chains of a BS is $M$ whereas it is $k$ for the users. Therefore, at any time, only a maximum $M$ of these beams can be active at a BS, and $k$ beams at the users. }
%
The BS then serves these users in the same beam in a time-sharing manner. 
We assume that each BS has a total transmission power of $P_{tot}$ Watts and this power is equally divided among beams considering a maximum of $M$ active beams. Hence, for each beam, the transmission power is set as $P_{tx}=P_{tot}/M$. Note that while adaptive transmission power schemes could be applied depending on the number of active beams, for simplicity, we assume a fixed transmission power per beam, i.e., $P_{tx}$. 

The antenna gain is denoted by $g^b(\alpha_{ji})$ for the transmitter of the link $ij$ and by $g^u(\alpha_{ij})$ for the receiver of this link.
Following the mmWave-based IEEE 802.15.3c standard \cite{toyoda2011antenna}, this gain in dB is defined as follows:
 \begin{align}
   g^b(\alpha_{ji}) &{=} \begin{cases}-0.4111\ln(\theta_{3\text{dB}}) {-} 10.579, \hspace{0.2cm} \text{   if } |\alpha_{ji}| {>} \frac{\theta^b}{2}, \\ 
   20 \log\left(\frac{1.6162}{\sin(\theta_{3\text{dB}}/2)}\right) {-} 3.01 \left(\frac{2\alpha_{ji}}{\theta_{3\text{dB}}}\right)^2, \text{ else,}\end{cases}
\end{align}
where $\alpha_{ij} = |\phi_{ij} - \zeta_{ij}|$ is the misalignment between the beam (boresight angle) and the geographical angle of link $ij$ in degrees, and $\theta_{3\text{dB}} = \theta/2.58$ is the 3 dB beamwidth. \newInfo{In our setting, other beam gain models can be considered easily without changing the model and proposed solutions in the following sections. }

\subsection{Channel model}
 \textcolor{black}{Similar to prior studies such as \cite{petrov2017dynamic, sopin2022user}}, we use the 3GPP path loss model~\cite{3gpp38901} for mmWave networks that depends on the 3D distance between user $i$ and BS $j$, denoted by $r_{ij}$. Since mmWave links are easily blocked by buildings, trees, or people, we distinguish links that are in line-of-sight~(los) and in non-line-of-sight~(nlos). The path loss in dB is: 
\begin{align}
    \ell_{\text{los}}(r_{ij})&= 32.4 + 21\log_{10}(r_{ij}) + 20 \log_{10}(f_c) + \text{SF}_{\text{los}}, \label{eq:los_path_loss}\\
    \ell_{\text{nlos}}(r_{ij}) &=\max \big\{ \ell_{\text{los}}(r_{ij}),  35.3 \log_{10}(r_{ij}) + 22.4\nonumber \\ &\hspace{1.5cm} + 21.3\log_{10}(f_c) + \text{SF}_{\text{nlos}}\big\}.\label{eq:nlos_path_loss}
\end{align}
Here, $f_c$ is the centre frequency in GHz, and $\text{SF}_{\text{los}}$ and $\text{SF}_{\text{nlos}}$ are standard normally distributed random variables with standard deviation $4$ and $7.82$ dB, respectively. To calculate the 3D distance, we assume the height difference between a user and a BS is $22.5$m \cite{3gpp38901}. In the optimization model that is introduced in Section \ref{sec:optimization}, we assume \newInfo{links are in line-of-sight or not based on the line-of-sight probability $p_\text{los}$ as given in \cite{3gpp38901}}:
\begin{align}
    p_\text{los} = \begin{cases}
    1, &\hfill r_{ij}^{\text{2D}} \leq 18\,\text{m},\\
    \frac{18}{r_{ij}^{\text{2D}}} +\left(1 - \frac{18}{r_{ij}^{\text{2D}}} \right) e^{-\frac{r_{ij}^{\text{2D}}}{36}}, &r_{ij}^{\text{2D}} > 18\,\text{m},
    \end{cases}
\end{align}
\newInfo{where $r_{ij}^{\text{2D}}$ denotes the 2D distance between a user $i$ and BS $j$.}
In Section~\ref{sec:resilience}, we also consider non-line-of-sight due to simulated blockers in the environment.

\subsection{Spectrum access scheme and associated overhead}

{At the beginning of every time slot, each user decides to which BS(s) to connect by following the UA scheme of this network. During the beam sweeping phase, users measure the surrounding BSs and send their connection requests for the BS with best signal through separate control channels. BSs receiving connection requests accept association requests based on whether a beam is available. Consequently, users direct their beam(s) towards the chosen BSs.}

When users have multiple links, we assume that the core network splits the traffic across all links, which means that the total capacity per user is the sum over all link capacities~(\textit{load balancing} \cite{Suer2020Multi-ConnectivityOverview}). We denote by $\B^i$ the set of BSs to which user $i$ connects.  To account for the overhead that is caused by load balancing and beam management \cite{aghashahi2021stochastic}, {we introduce the overhead factor 
$\xi$ which represents the time lost for beam sweeping and load balancing procedures. Consequently, the actual perceived capacity is a fraction $(1 - \xi) \leq 1$ of the total link capacity. 
{This overhead factor depends on, among other factors, the beamwidth: the smaller the beamwidth, the longer the beam sweeping time is. Therefore, small beamwidth results in higher complexity than larger beamwidths~\cite{paul2020boost, gao2020deep}. }
{ In addition, a smaller beamwidth leads to higher hardware complexity as it requires more antenna elements to fully cover the area.}
{While finding the optimal beamwidth that minimizes overhead while keeping high signal quality~(e.g. \cite{9500381, perfecto2016beamwidth}) is an interesting research question, we will leave this for future work and assume a constant overhead factor in this study.} 

{Lastly, users who are close to each other might decide to connect to the same beam of the same BS. In such cases where the BS needs to serve multiple users in the same beam, we assume that the BS applies time-division multiple access~(TDMA) to serve all of these users sharing a beam.}

\subsection{SNR and channel capacity}
Because dense mmWave networks are rather noise-limited than interference-limited \cite{elshaer2016downlink, singh2015tractable}, we omit the interference in our user association problem for tractability purposes.

Now, let us first calculate the signal-to-noise ratio $\gamma_{ij}$ for link $ij$ as follows:
\begin{align}
    \gamma_{ij} = \frac{P_{\text{tx}} g^b(\alpha_{ji}) g^u(\alpha_{ij})}{N \cdot \ell(r_{ij})},
\end{align}
in which the noise $N$ is the total noise over the operation bandwidth and equals to the sum of the thermal noise $N_0 $ and the noise factor $NF$. 
{In Section \ref{sec:simulations} we provide an analysis of channel capacity and resulting satisfaction considering the interference in the network with SINR  defined as:}
\begin{align}
    \gamma_{ij}^{+} = \frac{P_{\text{tx}} g^b(\alpha_{ji}) g^u(\alpha_{ij}) \ell(r_{ij})^{-1}}{N  + \sum_{k\in \B, k\neq j} P_{\text{tx}}g^b(\alpha_{ki})g^u(\alpha_{ik}) \ell(r_{ik})^{-1}s^{d_{ik}}_k},
\end{align}
{where $s^{d_{ij}}_j$ equals $1$ if the beam of BS $j$ in beam $d_{ij}$ (the beam towards user $i$) is active and $0$ when it is not active.}


We assume that a link must maintain a minimum SNR level $\gamma_{\textrm{min}}$ to ensure the decodability of the received signal. Consequently, the Shannon's channel capacity for user $i$ denoted by $C_{i}$ can be defined as follows:
\begin{align}
    C_{i} &= (1 - \xi) \sum_{j \in \B^i} C_{ij} \nonumber \\
    &= (1 - \xi) \sum_{j \in \B^i} x_{ij}  W  \log_2(1 + \gamma_{ij}),\label{eq:channel_capacity}
\end{align}
where the factor $x_{ij} \in [0, 1]$ denotes the fraction of time BS $j$ serves user $i$ in this beam. When $x_{ij} < 1$, this means that the beam is shared with other users. 

Given that the user's application requires at least $R_{\text{min}}$ bps,  we define the satisfaction $p_i$ of user $i$ as the ratio of the maintained data rate over the required minimum rate $R_{\text{min}}$. More formally,  
\begin{align}
    p_i = \min\left\{1, \frac{C_i}{R_{\text{min}}}\right\}.\label{eq:satisfaction}
\end{align}

\section{Optimal User Association with MC}
\label{sec:optimization}
In this section, we formulate an optimization model to find the optimal UA scheme that supports MC. 
{Let the decision variable $x_{ij} \in [0,1]$ denote the fraction of time the link between user $i \in \U$ and BS $j \in \B$ is used. For example, $x_{ij} = 0$ indicates that there exists no connection between user $i \in \U$ and BS $j \in \B$, and $x_{ij} = 1/2$ represents a scenario where BS $j$ has two connected and assigns half of its total service time to both users.} %

Since a cellular user's satisfaction is heavily influenced by the achieved downlink data rate, we consider a network operator that aims at maximizing its total network throughput while ensuring that users can maintain downlink rates above their minimum required rate. 
We represent the total throughput in the first term of \eqref{P1}. 
While the network operator needs to meet the minimum rate requirements of its users, this might not be possible under certain cases, e.g., heavy traffic load or high user density. 
To account for such cases where there is no feasible solution that meets all rate requirements, we introduce a penalty $Z$ for each user whose rate requirements are not met and thereby is \textit{unsatisfied}.
This penalty factor is represented in the second term of \eqref{P1}.
Depending on the network operator's policy, penalty $Z$ can be tuned to have a balance between the achieved network throughput and the number of unsatisfied users. 
For example, to provide an equal trade-off between connectivity and throughput, the value of $Z$ should be approximately the average channel capacity of a user.
Higher values of $Z$ would prioritise a (weak) connection over high throughput.

Now, let us introduce the optimization problem: 
\begin{align}
    \max_{\mathbf{x}} \sum_{i\in \U, j\in \B}(1 {-}\xi)x_{ij}& W  \log_2\left(1{+} \gamma_{ij}\right) {-} Z\sum_{i \in \U}(1 {-} p_i),  \label{P1}
\end{align}
subject to:
\begin{IEEEeqnarray}{rlll}
    s_i^d &\leq& 1, &\forall i \in \U, d \in \D^u, \IEEEyessubnumber\label{C2}\\
    \sum_{d\in \D^b}s_j^d &\leq& M, &\forall j \in \B, \IEEEyessubnumber\label{C2a}\\
    \gamma_{ij} &\geq& \one_{\left(x_{ij}\geq 1\right)} \gamma_{\text{min}}, &\forall i \in \U, j \in \B, \IEEEyessubnumber\label{C3}\\
    \sum_{j \in \B} \one_{\left(x_{ij}\geq 1\right)} &\leq& k &\forall i \in \U, \IEEEyessubnumber \label{C4a}\\
    \sum_{j \in \B} Wx_{ij} \log_2&(1& + \gamma_{ij}) \geq p_i R_{\text{min}}, \hspace{0.1cm}&\forall i \in \U, \IEEEyessubnumber\label{C4}\\
    x_{ij} &\in& [0,1], &\forall i \in \U, j \in \B, \IEEEyessubnumber\label{C5}\\
    p_{i} &\in& [0, 1], &\forall i \in \U. \IEEEyessubnumber\label{C6}
\end{IEEEeqnarray}

Constraint \eqref{C2} ensures that every beam $s_i^d$ of user $i$ can be used only once. Constraint \eqref{C2a} ensures that per BS, at most $M$ beams can be active at the same time reflecting the hardware limitation of the BS due to its $M$ RF chains. 
In Constraint \eqref{C3}, the indicator function $\one_{\left(x_{ij} > 0\right)}$ yields 1 if $x_{ij}$ is nonzero and this constraint ensures a minimum SNR for every existing link. Similarly, Constraint \eqref{C4a} ensures that there are at most $k$ connections per user. For single-connectivity, we set $k = 1$. Constraint \eqref{C4} states that the achieved user throughput must be at least the minimum rate requirement $R_{\text{min}}$ multiplied by the decision variable $p_i \in [0,1]$, which is equal to $1$ if user $i$ maintains its minimum rate requirement and can be less than 1 if there is no feasible solution meeting all users' rate requirements. \newInfo{Moreover, due $x_{ij}$ being a fraction, this constraint takes into account that the transmission time from BS $j$ might be split over multiple users.} Constraint (\ref{C5}) and (\ref{C6}) denote the types of the decision variables. 

The optimization model in \eqref{P1} is a linear mixed-integer program that is typically hard to solve optimally in real-time for large, realistic settings (the problem is similar to a knapsack problem, which is NP-hard to solve \cite{jin2022toward, chen2020impact}) \newInfo{and when it is possible to solve these optimization problems optimally, often only for small settings \cite{chen2022joint}}. We therefore first solve the formulated problem with Gurobi optimization software \cite{gurobi} under various settings in the following section, and then design our low complexity UA scheme by drawing insights from these optimal outcomes. 

\section{\color{black} Numerical investigation of the optimal UA}
\label{sec:simulations}
In this section, we assess the behaviour of the optimal scheme via simulations to address the following questions:
\begin{itemize}
    \item What is the impact of the BS transmission beamwidth on the per-user throughput, satisfaction level, and the optimal number of connections~(i.e., the \textit{optimal degree of MC})?
    \item {What is the benefit of MC in terms of per-user throughput and satisfaction level? Under MC, how many beams are active? } 
    \item What is the resulting misalignment between a user's receiving antenna and its serving BSs' transmission beam under the optimal association scheme?
\end{itemize}

To address these questions, we simulate a hexagonal BS deployment on a grid of $800$m $\times 1040$m, with an inter-site distance $200$m, which results in $24$ BSs in total. Users are simulated by a Poisson point process with user density $\lambda_U$. The simulation parameters are given in Table \ref{tab:used_values}. To mitigate boundary effects, we use toroidal boundary conditions. \newInfo{For each possible user-BS link in the simulation, we choose either the line-of-sight path loss $\ell_{\text{los}}$ given in \eqref{eq:los_path_loss} with probability $p_\text{los}$ and $\ell_{\text{nlos}}$ otherwise. }

\begin{table}[]
\caption{Simulation parameters.}
\label{tab:used_values}
\centering
\begin{tabular}{|l|l|l|}
\hline
$\lambda_U$                  & $\#$users per  km$^2$           & $\{50, 100, 250, 500, 750\}$ \\
$\theta^u$, $\theta^b$ & beamwidth of user, BS         & $\{5, 10, 15\}$ degrees \\
$k$ & 
{maximum $\#$connections per user} & $\{1, 2, \infty\}$\\
$M$ & {maximum $\#$active beams} & {10 \cite{paul2020boost}}\\
$Z$ & penalty factor & $750$ \\
$f_c$           & carrier frequency                    & $28$ GHz            \\
$W$                   & bandwidth           &$200$ MHz\\
$P_{\text{tx}}$        & transmission power {per beam}           & $20$ dBm                 \\
$N_0$               & noise power & $-84$ dBm \cite{halperin2013simplifying}\\ 
$NF$ & noise factor & $7.8$ dB\\
$\gamma_{\text{min}}$  & minimum S(I)NR                  & $5$ dB   \\
$R_{\text{min}}$  & minimum rate                & 100 Mbps  \cite{series2017minimum}\\
$\xi$ & overhead factor & 0.25 \\ \hline
\end{tabular}
\end{table}

\begin{figure*}[h!]
    \subcaptionbox{Channel capacity.\label{sfig:capacity_beamwidth}}{\includegraphics[width=.33\textwidth]{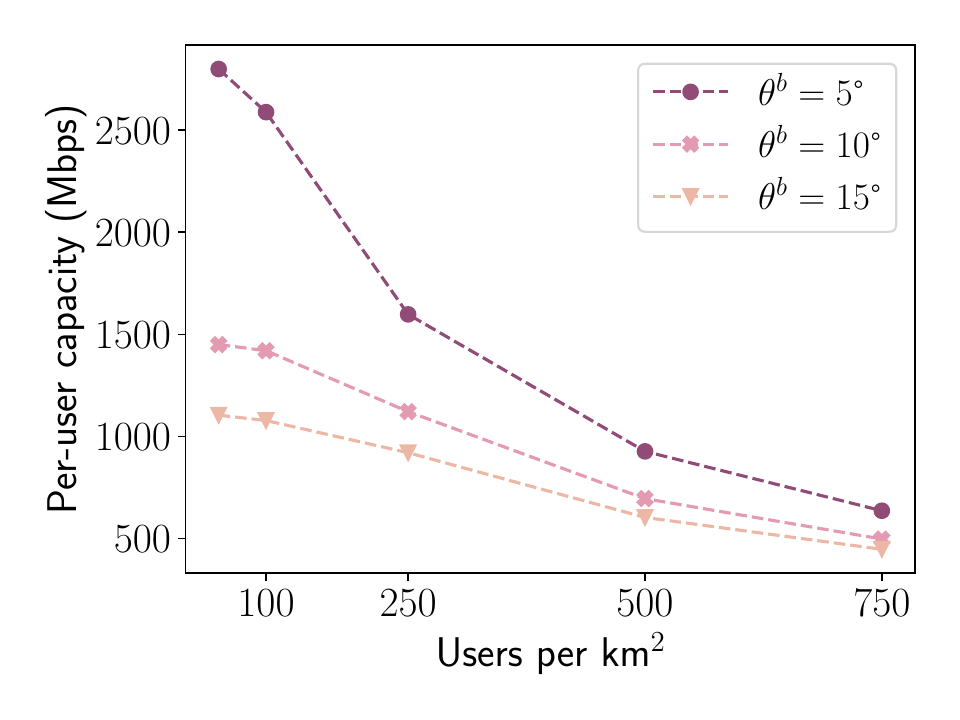}}\hfill
    \subcaptionbox{Satisfaction level.\label{sfig:disconnected_beamwidth}}{\includegraphics[width=.33\textwidth]{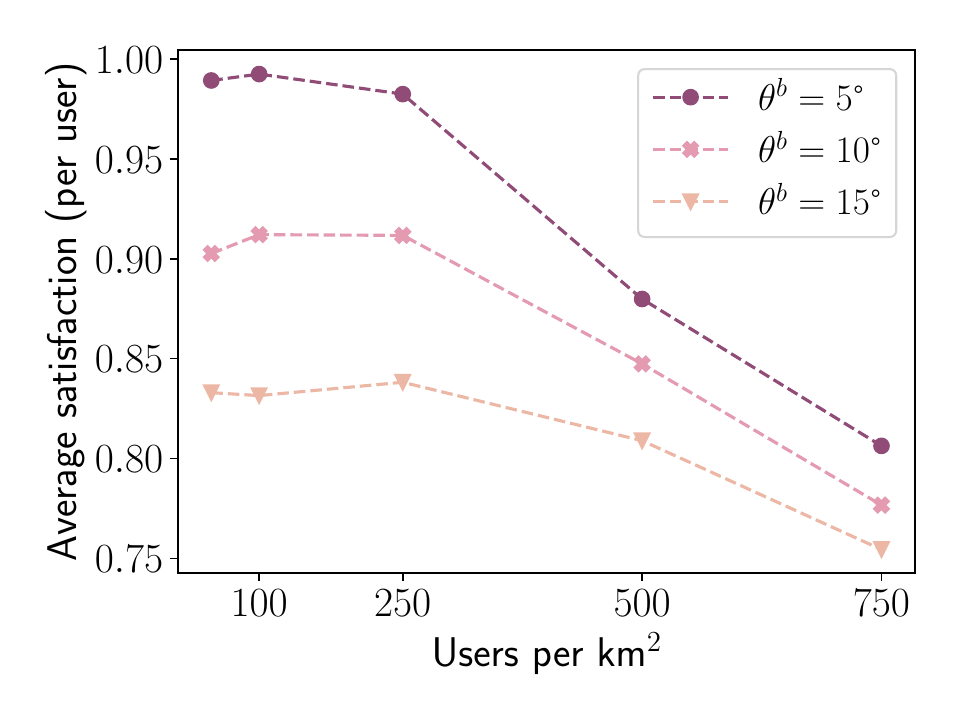}}\hfill
    \subcaptionbox{Number of connections per user.\label{sfig:degrees_beamwidth}}{\includegraphics[width=.33\textwidth]{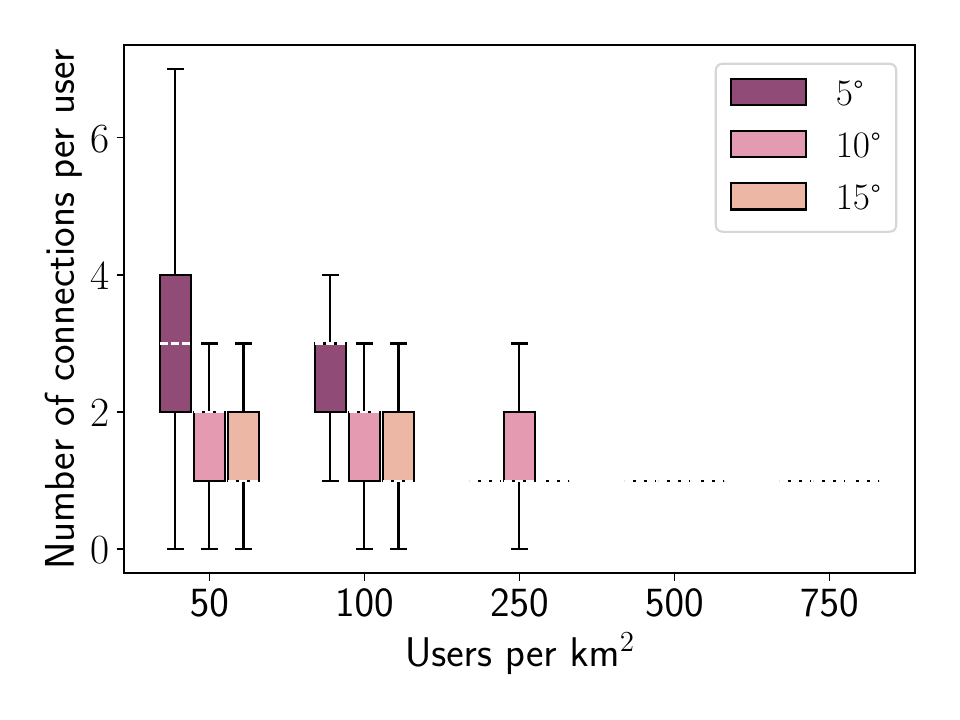}}\hfill
    \caption{Capacity, satisfaction level, and distribution of the number of connections under different beamwidths, $k = \infty$.} 
    \label{fig:different_beamwidths}
\end{figure*}

\begin{figure*}[h!]
    \subcaptionbox{Channel capacity.\label{sfig:capacity_mc}}{\includegraphics[width=.33\textwidth]{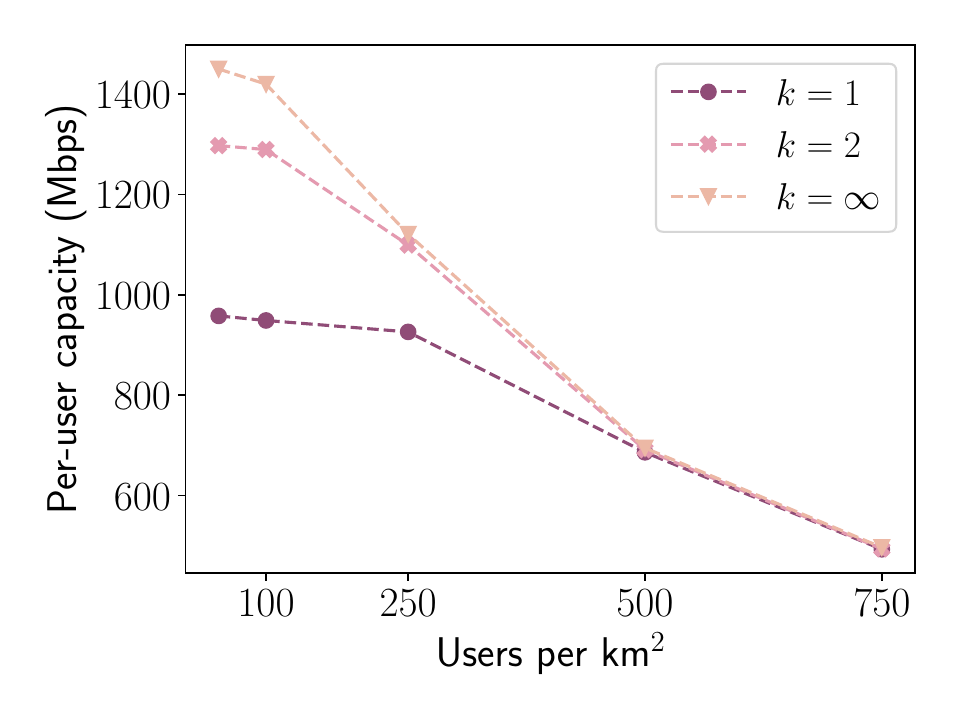}}\hfill
    \subcaptionbox{Satisfaction level.\label{sfig:disconnected_mc}}{\includegraphics[width=.33\textwidth]{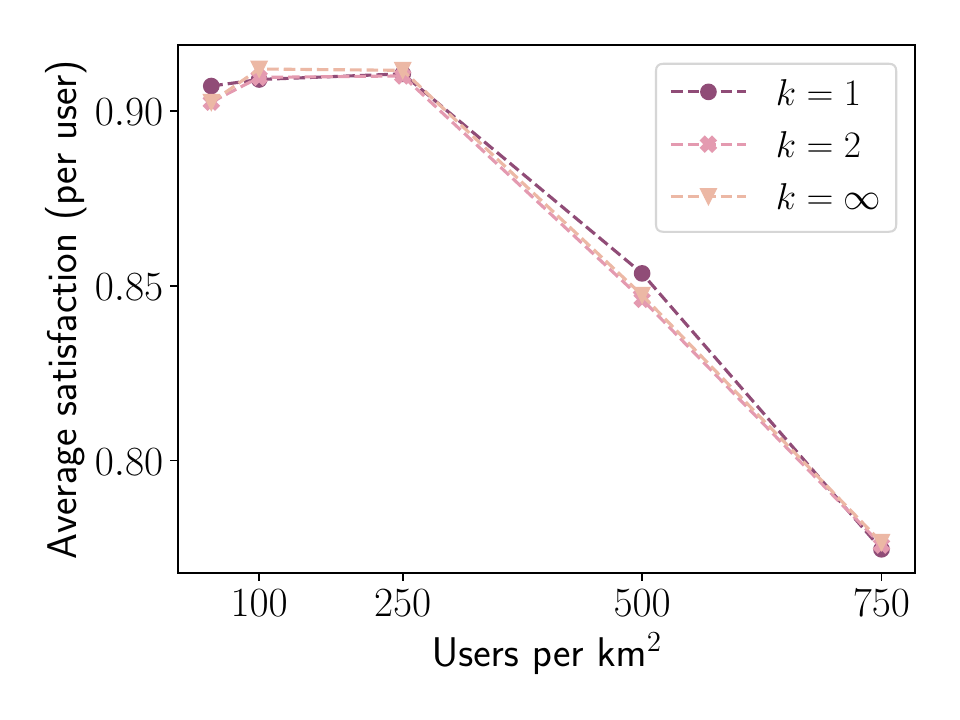}}\hfill
    \subcaptionbox{Number of active beams. 
    \label{sfig:active_beams}}{\includegraphics[width=.33\textwidth]{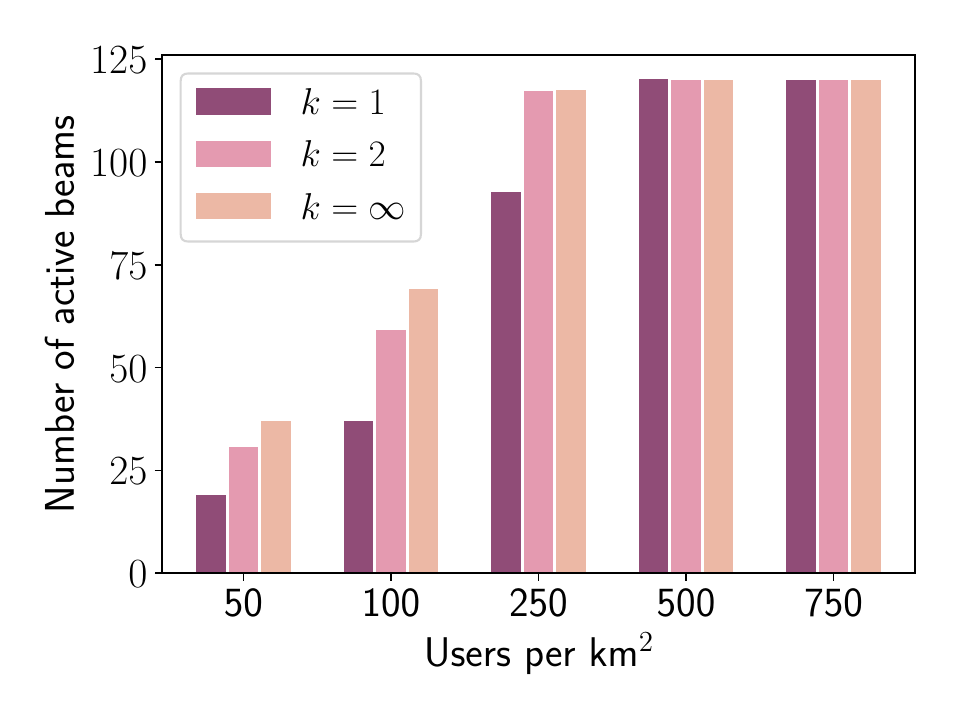}}\hfill
    \caption{Capacity, satisfaction level and number of active beams under SC and MC, $\theta^b = 10\degree$. } 
    \label{fig:different_mc}
\end{figure*}

We computed the optimal UA scheme in several simulated scenarios with varying beamwidth and number of users per beam, with as many iterations as needed to obtain in total 10.000 users. 
First, we show the impact of different beamwidths and varying maximum number of users per beam. We use three metrics to compare different scenarios: average per-user capacity calculated using \eqref{eq:channel_capacity}, average satisfaction level calculated using \eqref{eq:satisfaction} and fraction of disconnected users.


\subsection{Impact of BS transmission beamwidth $\theta^b$}
Fig.~\ref{fig:different_beamwidths} shows the average per-user channel capacity and satisfaction level under an increasing number of users for three BS beamwidths of $5\degree$, $10 \degree$ and $15 \degree$ whereas we assume a $5 \degree$ user beamwidth. As Fig.~\ref{sfig:capacity_beamwidth} shows, narrower beams result in higher throughput for all user densities. For 50 users, operating with $5\degree$ beams yields a $2.5\times$ throughput improvement over $15 \degree$ beams. This performance gap decreases with higher user densities: for 750 users, there is a $1.5\times$ improvement in throughput for $5\degree$ beams compared to $15\degree$ beams.
%

\newInfo{As seen in Fig.~\ref{sfig:disconnected_beamwidth}, 
smaller beamwidths achieve higher satisfaction, which is in line with the average channel capacity as observed in Fig.~\ref{sfig:capacity_beamwidth}, although the difference becomes smaller for increasing user density.}  Fig.~\ref{sfig:degrees_beamwidth} plots the distribution of the number of connections per user for the considered beamwidths. 
From this figure, we observe that the number of connections per user decreases with increasing network load but does not differ significantly for different beamwidths. However, for increasing user density, the number of connections per user decreases, indicating that MC is not beneficial in dense user settings.

Moreover, the satisfaction level and the fraction of disconnected users are similar: users are almost never partially served. {For example, for $\lambda_U = 250$, $1.8\%, 8.8\%$ and $16.1\%$ of the users are disconnected for 5, 10$\degree$ and 15$\degree$ beamwidth respectively. This corresponds to a satisfaction level of $0.98, 0.91$ and $0.84$ in Fig.\ref{sfig:disconnected_beamwidth}.

%

In Fig.~\ref{fig:different_mc}, we further explore the differences between MC and SC. We provide three scenarios: either single-connectivity ($k =1$), MC with a maximum of two connections ($k = 2$) or $k =\infty$, which denotes an MC setting where users can have as many connections as possible\footnote{In practice $k = \infty$ is not possible and does not happen, but it allows us to analyse the limits of multi-connectivity.}. Fig.~\ref{sfig:capacity_mc} shows that MC outperforms SC in all lower user densities in terms of channel capacity, while the results for SC and MC are similar when $\lu \geq 500$. In this setting, the average number of connections per user is 1 in all scenarios, which means that the optimal solution does not use multi-connectivity. In terms of satisfaction, SC and MC achieve the same results, as observed in Fig.~\ref{sfig:disconnected_mc}. In sparse user scenarios, Fig.~\ref{sfig:active_beams} shows that under SC and MC with $k =2$ and $k = \infty$, not all beams are used. Thus, MC results in higher per-user capacity and higher energy usage, but on average the same energy efficiency (these results are omitted for the sake of brevity).}

In a nutshell, there is no one-size-fits-all optimal degree of MC; it depends on the user density and the number of available beams in the network and also varies per user, illustrating the complexity of the optimal UA scheme.

\subsection{Distribution of the misalignment angle $\alpha_{ij}$}
Fig.~\ref{fig:connect_to_bs} shows the probability of connecting to the red BS (denoted by a star) in the optimal scheme, averaged over 1000 runs with $\lambda_U = 750$. We rounded every location of the user in every simulation to the nearest integer value. The dark blue lines in this figure indicate that users in the middle of these beams have a high probability of connecting to that BS. Thus, for dynamic MC UA schemes, user locations and angular orientation are  key factors. 
\begin{figure}
    \centering
    \includegraphics[width = 0.4\textwidth]{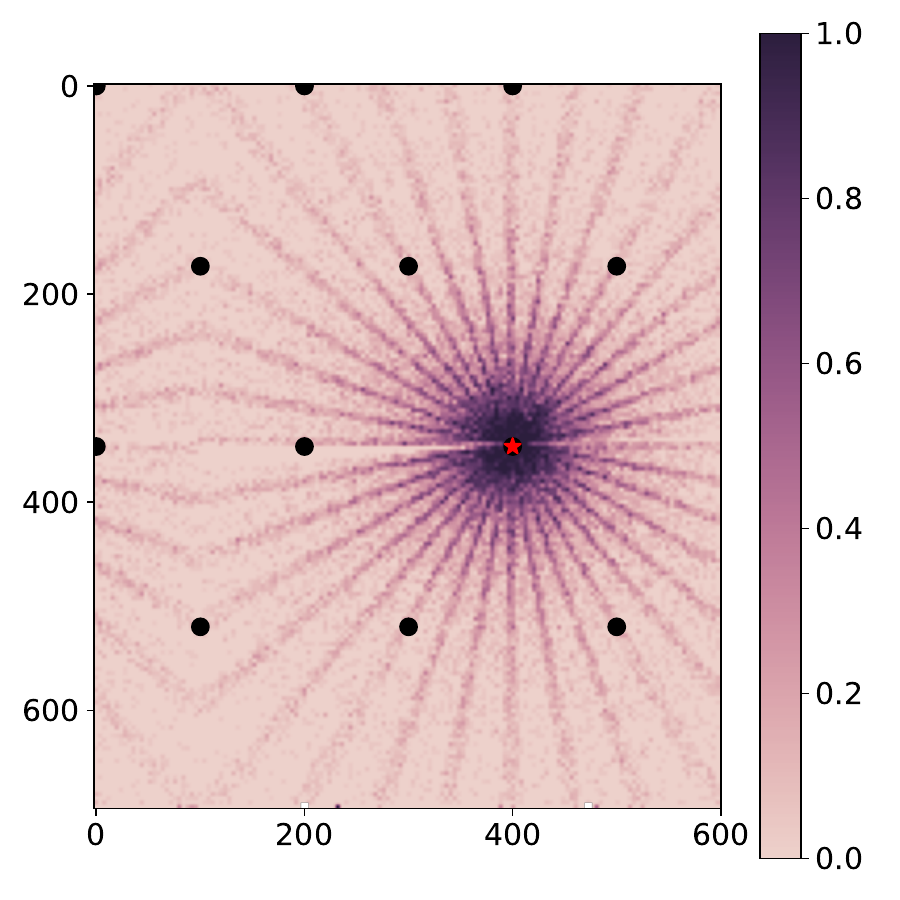}
    \caption{Probability of connecting to the red BS. The beamwidth is $\theta^b = 10\degree$, which means that the boresight angle of the beams are at $0, 10, 20, 30, \ldots$ degrees.}
    \label{fig:connect_to_bs}
\end{figure}

To investigate this observation quantitatively, we also plot in Fig.~\ref{fig:misalignment_750_users} a histogram of the misalignment of all existing links in the optimal UA scheme, i.e., the misalignment $\alpha_{ij}$ for all links $ij$ such that $x_{ij} = 1$. 
This figure has three \textit{peaks}: one main peak in the middle, where the misalignment from both the BS's side and the user's side is small; and two smaller peaks, where the user's misalignment is small. 
This distribution in misalignment angles corroborates our earlier observation that the angular orientation of a user is an important factor in the UA decision, as the misalignment distribution would be almost uniform if users connect to a BS disregarding the angular orientation with respect to the BSs. Moreover, Fig.~\ref{fig:connect_to_bs} shows that the distance from the user to the BS also plays a role in this decision, as it shows that the connection probability decreases when users are further away from the red BS. In fact, the path loss increases for longer distances. Therefore, correctly aligned users closer to the red BS have a higher probability of connecting to it compared to further away users that have a lower SNR and thus lower throughput.

\begin{figure}
    \centering
    \includegraphics[width = 0.35\textwidth]{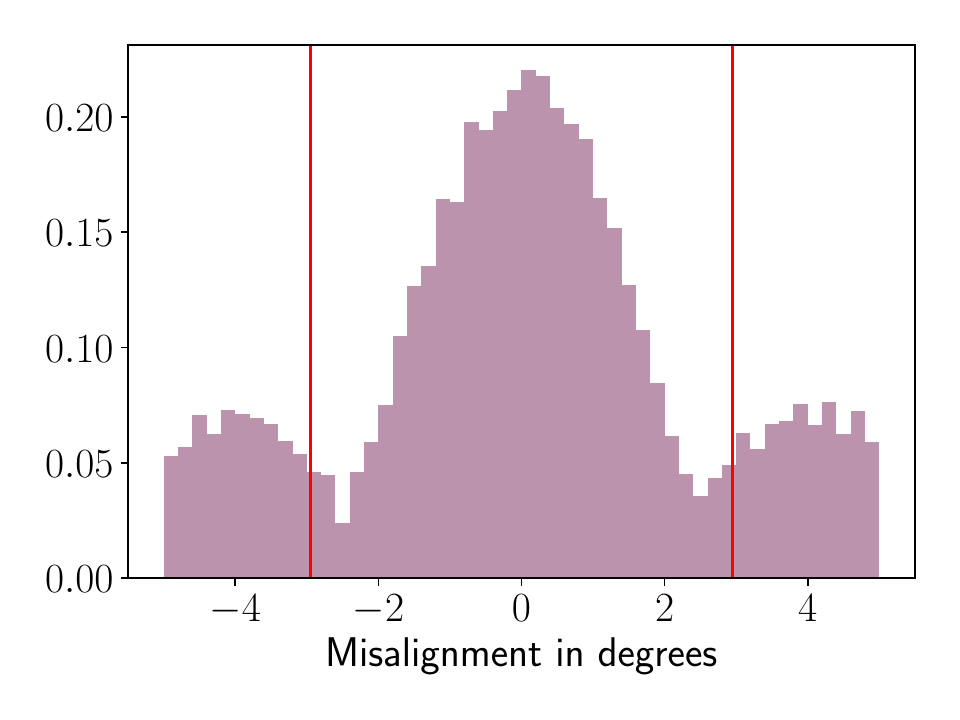}
    \caption{Histogram of the misalignment of all links in the optimal UA scheme for $\lambda_U = 250$, $\theta^b = 10\degree$ and $k = \infty$. The red line denotes the threshold $\sigma(\theta^b)$, which is twice the standard deviation of the misalignment.}
    \label{fig:misalignment_750_users}
\end{figure}

\section{BEAM-ALIGN: UA considering beam misalignment} 
While we devised the optimal UA scheme in the earlier sections, it is impractical due to the associated high computational complexity and the need for information exchange between the BSs and a central entity, e.g. controller. The optimization scheme is a multi-knapsack problem, which is known to be NP-hard.
Therefore, in this section, we design a near-optimal distributed UA scheme, which is only based on local user information, so that it will not involve any communication between BSs. Based only on the experienced received SINR from a BS, each user sends a connection request to this given BS. Then, the BS, given all connection requests, decides whether to connect or not. Consequently, the UA scheme can run efficiently and exhibits high scalability, which makes it favorable in real deployments with many BSs and users.  

Based on our observation that misalignment is a key factor in deciding whether to connect to a BS, we devise \textsc{beam-align} whose pseudocode is shown in Alg.~(\ref{alg:beamwidth_heuristic}).
In \textsc{beam-align}, a user connects to every BS for which the BS-user  misalignment is smaller than a threshold $\sigma(\theta^b)$, and if the number of active BS beams when accepting the request does not exceed $M$. When the number of incoming requests for a beam is higher than the maximum allowed $s$, the BS gives priority to users with high SINR to maximize the achieved throughput.
The time that a BS serves a user is split equally among the users who are in the same beam.
This algorithm can be run in a distributed manner at each BS without global knowledge of other BSs and users. For instance, at the beginning of a time slot, users can send their connection requests to the BSs through the uplink control channels and each BS can accept or reject the requests based on the misalignment~(Line~\ref{alg:ifmisalign}) and total number of received requests~(Line~\ref{alg:iftotalusers}). To maintain the minimum signal quality for decodability of the signal, users can only send a connection request to a BS if the link-SNR is larger than $\gamma_{\text{min}}$.
As the complexity of sorting the incoming requests according to the corresponding link SINRs~(in Line~\ref{alg:sortusers}) is dominant in comparison to deciding to accept a user's request~(in Line~\ref{alg:foreachuser}-Line~\ref{alg:addLink}), the complexity of running \textsc{beam-align} is $O(|\mathcal{U}|\log|\mathcal{U}|)$.
\begin{algorithm}[tb]
\begin{algorithmic}[1]
\State \textbf{input :} set of base stations $\B$ with known locations, set of users $\U^b$ that send a connection request to BS $b\in \B$.
\State \textbf{output:} $\mathcal{L}$: set of active user-BS links $l_{u, b}$ in the network.
\State beams$(b)$: number of active beams of BS $b$.
\\ 

\For{\textbf{each} $u \in \U$} \Comment{runs at every user distributedly}
\For{\textbf{each} $b \in \B$}
\If{SINR$_{u,b} \geq \gamma_{\text{min}}$ \textbf{and} $|\alpha_{bu}| < \sigma(\theta^b)$}
    \State{Send association request to BS $b$.}
\EndIf
\EndFor
\EndFor
\\ 
\State $\mathcal{L}= \{\}$   
\For{\textbf{each} $b \in \B$}\Comment{runs at each BS independently.}
\State \textit{Sort the incoming requests from $\U^b$ in descending order according to the link-SINR.}\label{alg:sortusers}
    \For{\textbf{each} $u \in \U^b$} \label{alg:foreachuser}
        \If{
        $|\alpha_{bu}| < \sigma(\theta^b)$} \label{alg:ifmisalign}
        \If{$|$beams$(b)$ + $\{s_u^{d_{ub}}\}| \leq M$:} \label{alg:iftotalusers}
        \State{$\mathcal{L}$ = \{$\mathcal{L}$,  $l_{u, b}$\}}  
        \State{beams$(b)$ {+=} $s_u^{d_{ub}}$ \label{alg:addLink}}
        \EndIf
        \EndIf
	\EndFor
\EndFor
\State{\Return $\mathcal{L}$}
\end{algorithmic}
\caption{\textsc{beam-align}: a distributed user association scheme that runs locally at every BS.}
\label{alg:beamwidth_heuristic}
\end{algorithm}

\begin{figure*}[thb]
\centering
    \subcaptionbox{Channel capacities.\label{sfig:HHO_capacity}}{\includegraphics[width=.24\textwidth]{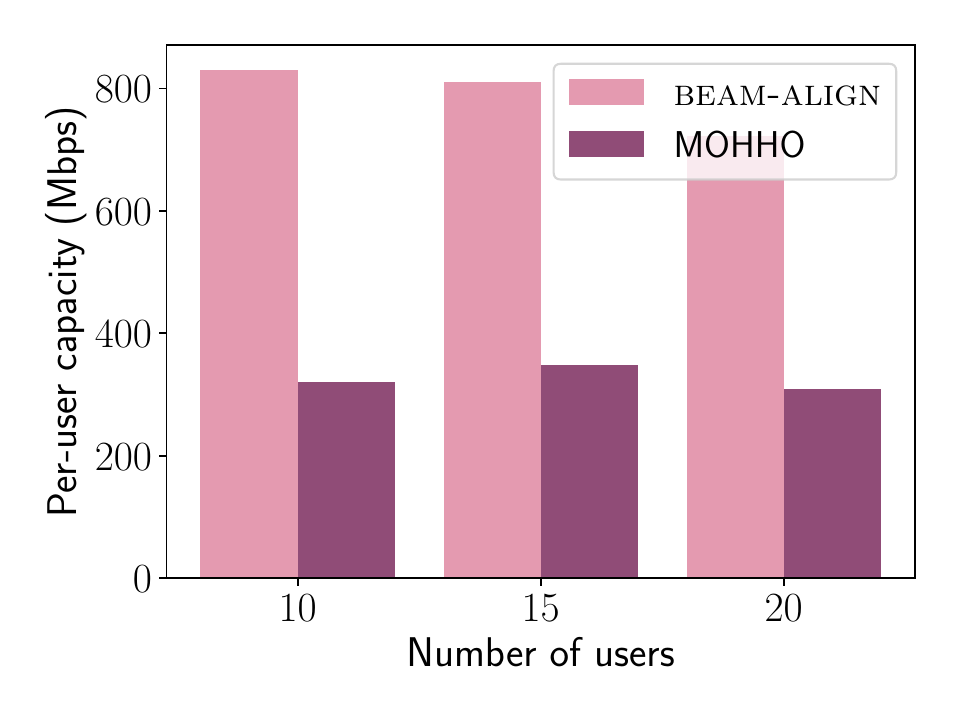}}\hfill
    \subcaptionbox{Satisfaction level.\label{sfig:HHO_satisfaction}}{\includegraphics[width=.24\textwidth]{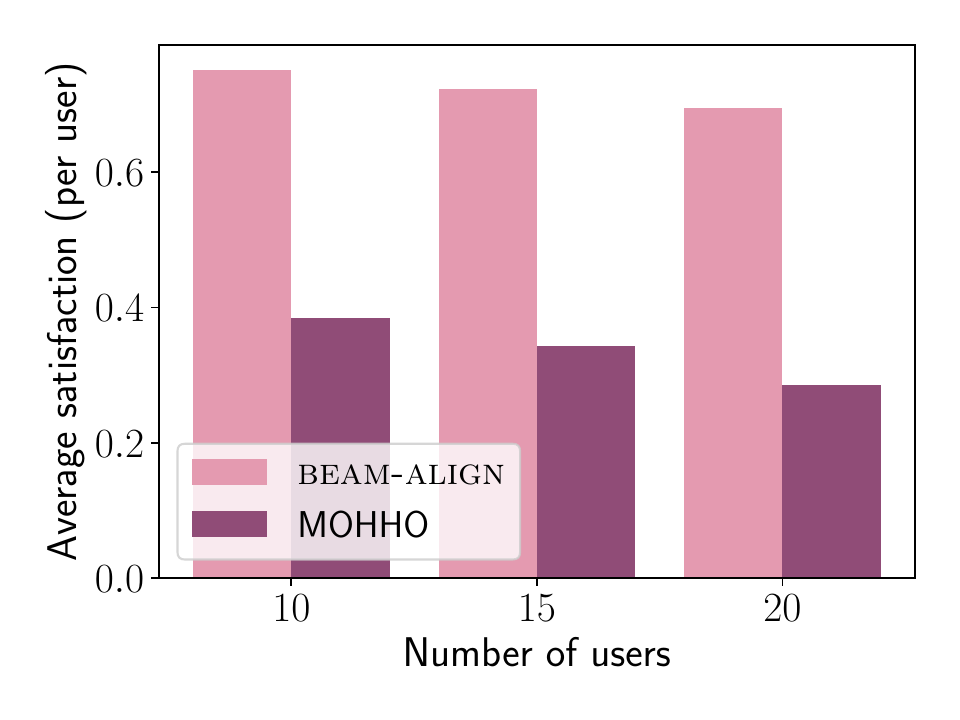}}\hfill
    \subcaptionbox{sd of user rates (solid) and BS loads (striped).\label{sfig:sigmaU}}{\includegraphics[width=.24\textwidth]{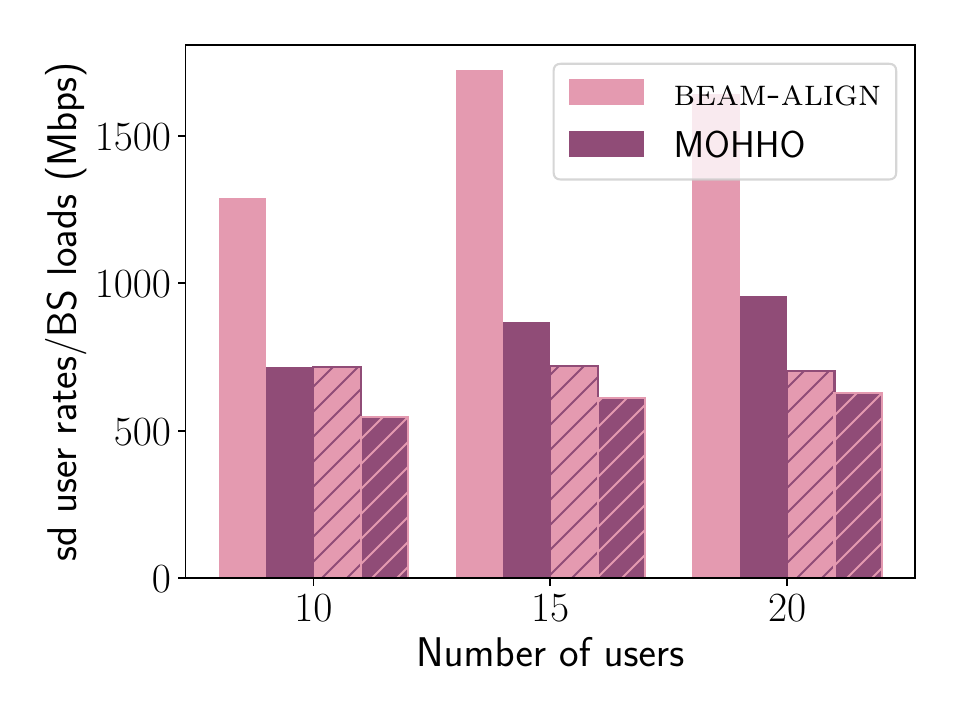}}\hfill
    \subcaptionbox{Energy efficiency. \label{sfig:sigmaU}}{\includegraphics[width=.24\textwidth]{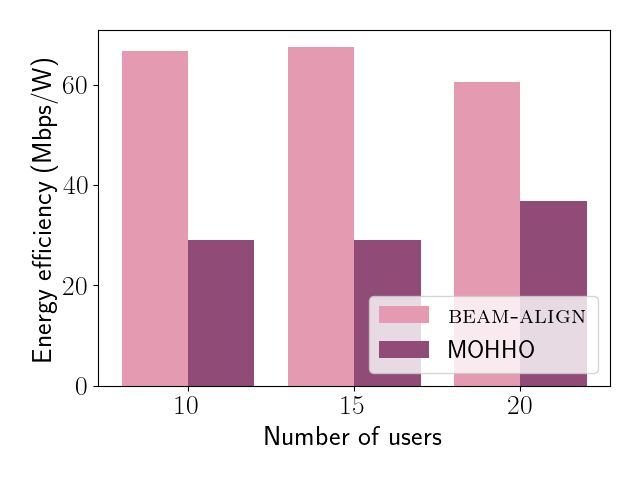}}\hfill
    \caption{Capacity and satisfaction level (based on SINR), and standard deviation (sd) of the user rates and BS loads, for \textsc{BEAM-ALIGN} and MOHHO, $\theta^b = 10\degree$ and $k = \infty$.} 
    \label{fig:HHO}
\end{figure*}

\begin{figure*}[thb]
\centering
    \subcaptionbox{Channel capacities.\label{sfig:capacity_heuristics}}{\includegraphics[width=.33\textwidth]{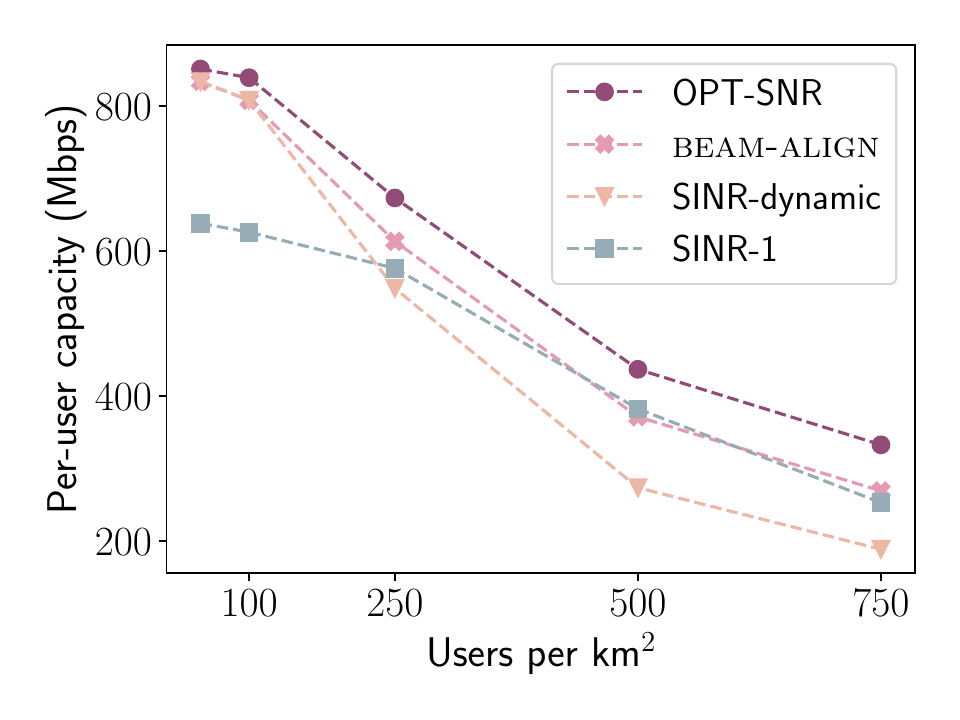}}\hfill
    \subcaptionbox{Satisfaction level.\label{sfig:satisfaction_heuristics}}{\includegraphics[width=.33\textwidth]{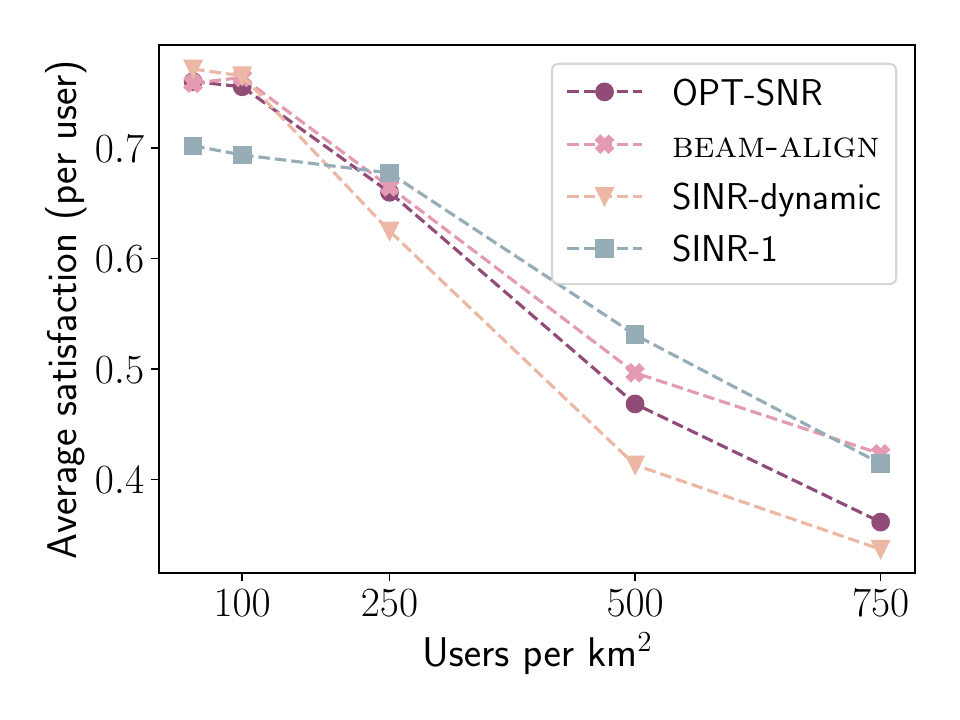}}\hfill
    \subcaptionbox{Number of connections per user.\label{sfig:degrees_heuristics}}{\includegraphics[width=.33\textwidth]{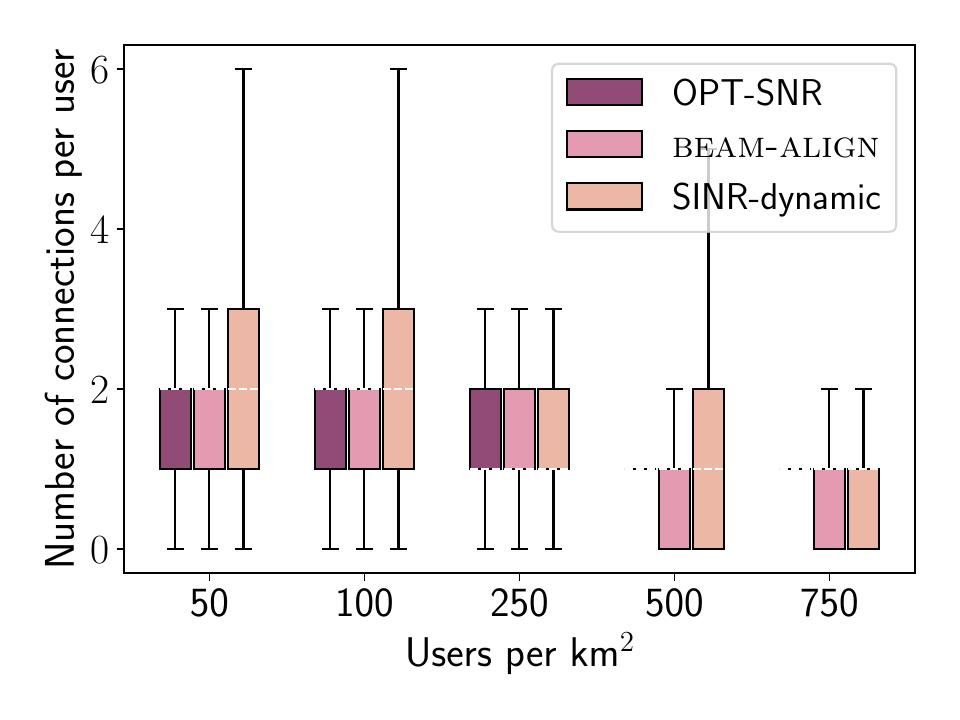}}\hfill
    \caption{Capacity and satisfaction level (based on SINR), and distribution of the number of connections, $\theta^b = 10\degree$. } 
    \label{fig:comparison_heuristics}
\end{figure*}

The misalignment threshold in Alg.\ref{alg:beamwidth_heuristic}~(Line~\ref{alg:ifmisalign}) is based on simulations of the optimization scheme and is equal to twice the standard deviation of the BS-user misalignment of all links in the optimal scheme. Note that this value is driven from our numerical analysis and can be tuned according to the preferences of the network operator. A smaller misalignment threshold will lead to higher throughput, although the fraction of disconnected users will also be higher.
This threshold can be derived offline by running the optimal algorithm under the applicable operation regime (e.g., the user beamwidth and BS beamwidths). The resulting values can then be saved at each BS to use while running this distributed heuristic. 

As there is no prior study considering user association in a mmWave setting with dynamic MC,, we consider {the heuristic algorithm called Multi-Objective Harris Hawks Optimization (MOHHO) \cite{jin2022toward} based on Harris Hawkes Optimization \cite{heidari2019harris}. This algorithm provides a heuristic solution to a similar optimization problem to ours, but uses a different objective: instead of maximizing throughput, MOHHO maximizes energy efficiency and minimizes the standard deviation of user rates and BS loads. To provide a fair comparison between MOHHO and \textsc{beam-align}, we ran both algorithms on 10, 15 and 20 users and 4 BSs as the running time of MOHHO running time significantly increases in larger settings. As MOHHO does not provide a constraint on the number of active beams, we constrained the power per BS to be similar to the transmission power per beam times the number of active beams in our method. Moreover, we used a swarm of $100$ feasible solutions and took $500$ time steps per iteration of the MOHHO algorithm. We simulated both \textsc{BEAM-ALIGN} and MOHHO for a total of 2,000 users\footnote{We share our code publicly on \url{https://github.com/lweedage/beam-align}.}.}

{Fig.~\ref{fig:HHO} compares the performance of \textsc{beam-align} and MOHHO considering interference for both schemes. As expected, our heuristic performs best on channel capacity and satisfaction level while MOHHO achieves the lowest standard deviation on both BS loads and user rates. However, our algorithm significantly outperforms MOHHO in terms of energy efficiency as well, with an increase of up to $229\%$ compared to the energy efficiency of MOHHO. Moreover, \textsc{beam-align} is outperforms MOHHO in terms of computational complexity. 
}

Next to the comparison with a state-of-the-art heuristic algorithm like MOHHO, we also compared our algorithm to two simple baselines that both rely on SNR for UA: \emph{SINR-1} and \emph{SINR-dynamic}. SINR-1 corresponds to UA in an SC scenario where a user connects to the BS with the highest SINR as in some previous studies~\cite{Gapeyenko2019OnDeployments, mezzavilla2016mdp}, provided that a beam is available.
In \emph{SINR-dynamic}, users connect to a beam of a BS with a SINR higher than $\gamma_{\text{min}}$ in order from highest to lowest SINR, until the maximum number of active beams is reached. 

Although misalignment and SINR are closely related as the signal degrades when the misalignment is large, \textsc{beam-align} and SINR-dynamic differ from each other as follows: \textsc{beam-align} results in a fairer UA scheme, as edge users that are well aligned will be preferred over users closer to BS but misaligned. The SINR-dynamic algorithm on the other hand, always prefers nearby users over edge-users, as nearby users have typically higher SINR. 

Fig.~\ref{fig:comparison_heuristics} shows the performance of the heuristics and the optimal solution in terms of capacity and satisfaction under increasing user density. \newInfo{These results are based on SINR instead of SNR to reflect a realistic scenario. As the problem stated in \eqref{P1} does not consider interference, the association decisions of \eqref{P1} might differ from a setting under interference. We denote this scenario by OPT-SNR.} {For all user densities, \textsc{beam-align} performs well and close to OPT-SNR, with a $2.0{-}19.1\%$ difference in per-user capacity.}
\textsc{beam-align} also very closely follows the satisfaction of OPT-SNR with an even better satisfaction rate compared to optimal for high user densities (Fig.~\ref{sfig:satisfaction_heuristics}). Interestingly, while the SINR-dynamic outperforms all other schemes for $\lu = 50$, the SINR-dynamic scheme results in an even lower per-user capacity and satisfaction level than the other UA schemes when $\lu > 100$. Compared to OPT-SNR there is an average decrease in per-user capacity of $21\%$.
Lastly, the SINR-$1$ scheme provides a more conservative UA scheme which results in a higher satisfaction level and per-user capacity in the medium dense regime $(\lu = 500)$, but a lower per-user capacity in the sparse user regimes. 
%
However, SINR-1 outperforms SINR-dynamic in terms of per-user capacity for user densities higher than $\lambda_U = 100$. Together with the fact that the satisfaction level for SINR-1 is also higher compared to SINR-dynamic, this indicates that when resources are limited, we conclude that users only benefit from MC if this is done smartly. 

Fig.~\ref{sfig:degrees_heuristics} shows the average number of links for UA schemes with MC. The average number of links per user in \textsc{beam-align} is similar to the SNR-OPT scheme. The SINR-dynamic results in a higher number of connections per user. {However, although many users have a connection~(only up to $9\%$ of the users are disconnected under SINR-dynamic), they do not receive enough resources from the BS to be satisfied with their connection. To summarize, \textsc{beam-align} results in a UA scheme that is suited for both high and low user densities, and follows the optimal degree of MC as computed in OPT-SNR.} 

\section{Robustness Analysis}\label{sec:resilience}
In the previous sections, we have not accounted for possible imperfections that mmWave links might experience and consequently lead to low performance. 
Here, we consider three scenarios to investigate the \textit{robustness} of optimal, \textsc{beam-align} and SINR-dynamic schemes under non-ideal conditions: namely, under blockage, rain, and clustered users. 
%
In case of blockage and rain, link quality degrades as the path loss increases when a link is blocked or affected by rain. 
If the resulting link SNR is lower than $\gamma_{\text{min}}$, the received signal cannot be decoded at the receiver, resulting in a link failure. Therefore, we recalculate the average per-user throughput and satisfaction level in these scenarios and compare this to the scenario without blockage or rain. We refer to this baseline as \textit{Normal}. 
To reflect more realistic user distributions rather than a uniform distribution, we consider a clustered user distribution. 
Next, we elaborate on these three scenarios.
\begin{itemize}[leftmargin=*]
\item \underline{Scenario I - Blockage:} We simulate blockers according to the method described in \cite{bai2014analysis} and \cite{sopin2022user}: for every blocker, we randomly generate a center, length, width and angle uniformly from a certain range, which denotes a rectangular area in the plane that is blocked. This process is done iteratively until the total blocker area is 10\% of the considered area, \newInfo{resulting in a similar blocker density as investigated in \cite{sopin2022user}}.  
We assume that the blockers are higher than the link between a user and a BS, and therefore we ignore the height of the blockers. 
Based on this set-up with blockers and the given UA schemes, we determine whether a link is blocked (nlos) or not (los), resulting in the real path loss of that link as described in Eqs.\eqref{eq:los_path_loss} and \eqref{eq:nlos_path_loss}.
\smallskip
 \item \underline{Scenario II - Rain:} mmWave links might fail under heavy rain due to the increased path loss caused by additional attenuation. To investigate how our UA schemes will be affected by rain, we model the attenuation loss ($AT$) due to rain as in \cite{zhao2001analytic}:
\begin{align}
    AT &= \gamma_R \cdot r_{ij},\\
    \gamma_R &= k \cdot R^\alpha
\end{align}
where $k = 0.124$, $\alpha = 1.061$ \cite{zhang2015coverage} for $28$ GHz band and $R$ is the rain rate in mm/h. In our evaluation, we consider three rain intensities: 2.5 mm/h (light rain), 25 mm/h (heavy rain) and 150 mm/h (monsoon)~\cite{pi2011introduction}.
\smallskip
 \item \underline{Scenario III - Clustered Users:} As user mobility follows some social patterns, the resulting user location distribution resembles a clustered distribution more than a uniform distribution. To account for this fact, 
 we use a Mat\'ern cluster process to model the locations of the clustered users~\cite{chun2015modeling}. In our setting, this point process consists of a group of 10 \textit{parent} points that are simulated by a Poisson point process. Around these \textit{parents}, we then generate a group of $|\U|/10$ \textit{offspring} points, uniformly distributed within a radius of $50$\,m reflecting the coverage radius of a mmWave transmitter. Contrary to the other two scenarios, where we investigate the existing UA scheme with increased path loss, in this scenario, we simulate clustered users as a new scenario.
\end{itemize}

\lotte{Fig.~\ref{fig:blockers_results} shows the average per-user capacity~(Fig.\ref{sfig:blockers_capacity}) and satisfaction level~(Fig.\ref{sfig:blockers_satisfaction}) for all studied UA schemes. 
As we can see in Fig.\ref{sfig:blockers_capacity}, \textsc{beam-align} performs under imperfect environment closer to optimal than the SINR-dynamic user association scheme does. The optimality gap varies between $4.3-9.7\%$ in terms of average per-user capacity, except in the blockage scenario, where \textsc{beam-align} performs worse by $65\%$. However, while the average per-user capacity for \textsc{beam-align} drops significantly in the blockage scenario, the satisfaction level only shows a decrease of $18\%$, which is similar to the rain scenarios. For the blockage scenario, this difference between OPT-SNR and the two heuristics can be attributed to the fact that the OPT-SNR is a blockage-aware optimization scheme, thus already accounts for the possibility of blockage and increased path loss, while \textsc{beam-align} and SINR-dynamic do not. To mitigate this performance loss for the heuristics, UA schemes can proactively consider the appearance of blockers and determine user associations based on possible loss of line-of-sight.}
\\ 
\lotte{Comparing Fig.~\ref{sfig:blockers_capacity} to Fig.~\ref{sfig:blockers_satisfaction}, we see a difference between the per-user capacity and satisfaction level of \textsc{beam-align} and SINR-dynamic for the rain scenarios. Since both SINR-dynamic and \textsc{beam-align} allocate users to BSs based on SINR~(and beam alignment), users that are further away or are very misaligned will not connect to a BS. When path loss increases even more, these far-away users might not be able to have a connection any more, as the heuristics prefer users that are well-aligned and/or have high SINR, which results in the lower satisfaction level. When comparing SINR-dynamic with \textsc{beam-align}, for all rain scenarios, SINR-dynamic performs around $10\%$ worse. For the clustered and blockage scenario, SINR-dynamic has almost the same results as with \textsc{beam-align}, both in terms of satisfaction as in capacity. These observations are comparable to the normal setting as shown in Fig.~\ref{fig:comparison_heuristics}. Moreover, we see a similar number of disconnected and non-satisfied users, with only $3-4\%$ of users being partially satisfied.}

\lotte{Interestingly, while per-user capacity in the clustered scenario is similar for all three UA schemes, the satisfaction remains higher for SINR-dynamic and \textsc{beam-align}, indicating more fairness in the distribution of per-user capacities in the heuristics. A reason for this is that the interference in a clustered scenario will be higher as well, which results in a lower experienced throughput for the users in the OPT-SNR UA. As \textsc{beam-align} already accounts for the interference and is more spatially aware due to the misalignment criterion, this heuristic achieves a higher satisfaction level.}

\begin{figure}[thb]
\centering
    \subcaptionbox{Average per-user capacity.\label{sfig:blockers_capacity}}{\includegraphics[width=.4\textwidth]{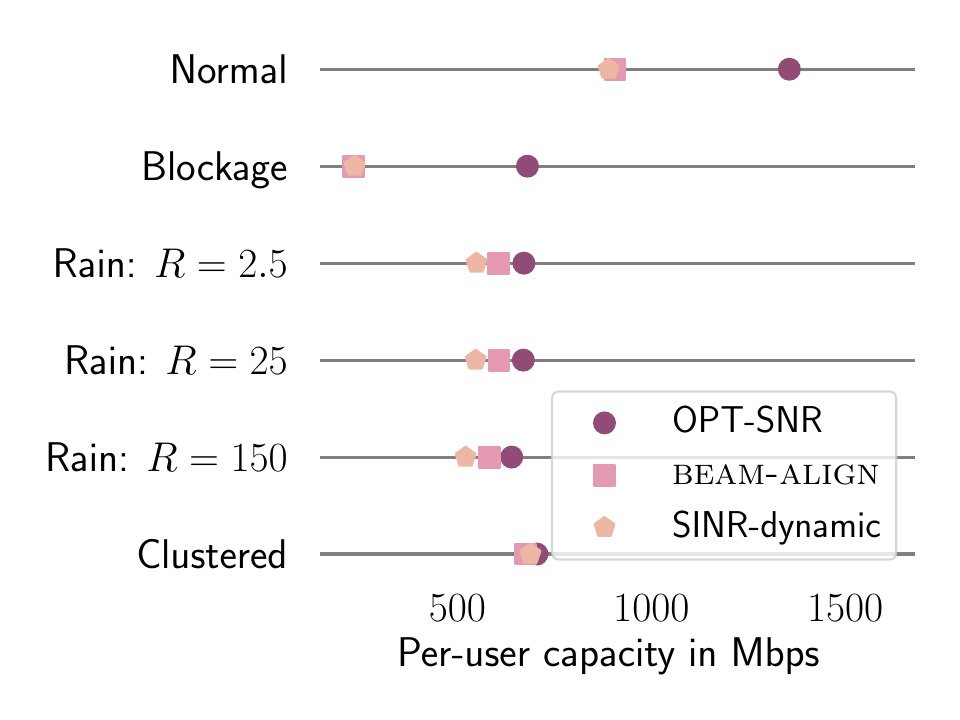}}\hfill
    \subcaptionbox{Average satisfaction level\label{sfig:blockers_satisfaction}. }{\includegraphics[width=.4\textwidth]{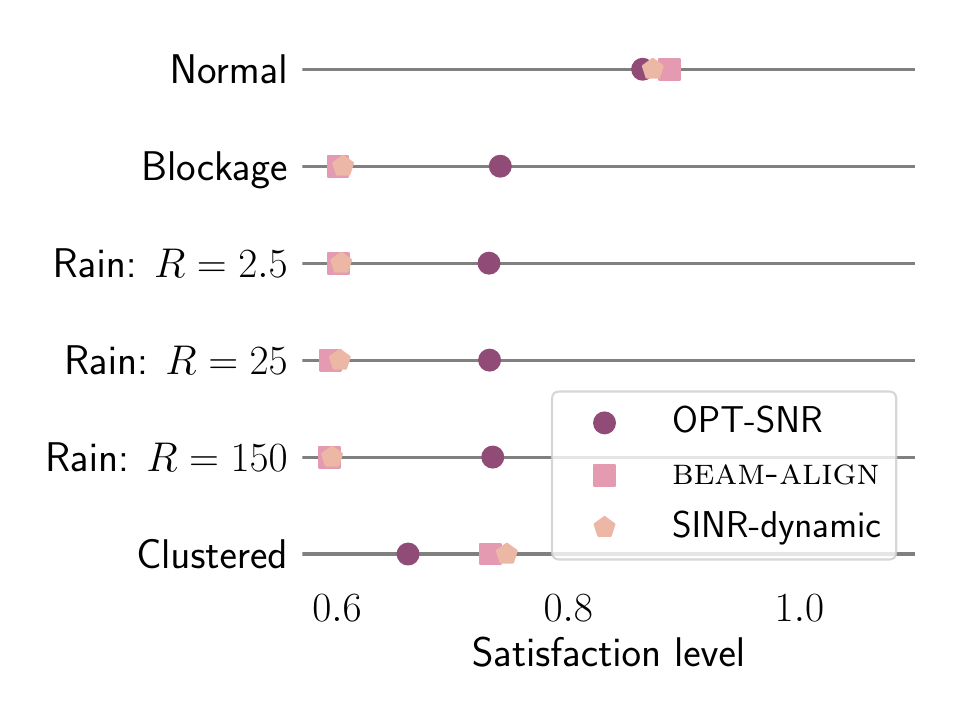}}\hfill
    \caption{Capacity and satisfaction level in the optimal scheme and two heuristics for three scenarios with blockage, rain with rate $R$ and clustered users, and $\lambda_U = 250$, $\theta^b = 10\degree$.} 
    \label{fig:blockers_results}
\end{figure}

\section{Discussion}\label{sec:discussion}
In this section, we discuss two shortcomings of our work: (i) limited consideration of overheads due to beamforming and MC and (ii) homogeneous user requirements. 

First, we considered a fixed time overhead~($\xi$) for {overheads due to MC and mmWave links, e.g., packet splitting overhead and beam alignment.} However, prior studies show that beam alignment overhead and the resulting performance degradation due to misalignment is larger for smaller beamwidths~\cite{haider2016mobility}. Hence, our conclusions derived from 
Fig.\ref{fig:different_beamwidths} which suggests that smaller beamwidths enable higher throughput may not hold if beamforming overhead is considerable.  %
Moreover, maintaining connections to the BS with smaller beamwidths becomes difficult for high mobility scenarios. 
Hence, considering an overhead factor that depends on the BS beamwidth is an interesting future research direction. Moreover, more sophisticated antenna operation can be considered, e.g., BSs can adapt their beamwidths depending on the network states, e.g., user density and mobility. 

Similarly, this work does not consider overheads due to MC such as scheduling control messaging. Consequently, under \textsc{beam-align}, some users are connected to as many as 12 BSs which might not be feasible in a realistic setting as the overhead will increase due to the inter-BS communication for packet splitting. Therefore, adding an extra constraint to the number of connections of a user in \eqref{P1} can limit the maximum number of connections to an upper bound. However, general trends observed in this paper are not expected to change, as longer-distance connections have lower spectral efficiency and therefore do not contribute much to the total throughput of the user.
Further investigation of the cost of MC and beam alignment can provide a more thorough understanding of the impact of MC and beamwidths.

The second limitation is due to our assumption of users with identical requirements. A realistic scenario can also include more heterogeneous rate or reliability requirements. 
To account for these heterogeneous requirements, we could expand \textsc{beam-align} by for example dividing the time in a beam heterogeneously depending on the user requirements or by ensuring a minimal number of connections for users requiring high reliability. Additionally, we assumed that BSs have identical antennas, e.g., beamwidths and fixed beam directions. We believe that further investigation in more dynamic antenna configurations can reveal even more benefits of MC and beamforming.


\section{Conclusion}\label{sec:conclusion}
Multi-connectivity can facilitate higher throughput and increased robustness against channel impairments frequent in mmWave networks. However, determining the optimal number of connections and user-BS association remains a challenge.  
In this paper, we formulated an optimization problem that solves the user association problem for mmWave networks with MC and beamforming, while maintaining a high user satisfaction level. Our numerical investigations showed that a smaller BS beamwidth results in a higher per-user capacity and a higher satisfaction level. 
Inspired by the observations on the optimal association scheme with SNR, we designed \textsc{beam-align} that only needs local information based on the misalignment between the geographical angle of a user-BS link and the boresight angle of the beam to which this user connects. We showed that \textsc{beam-align} performs close to optimal in terms of per-user capacity while still keeping a moderately high satisfaction level, when incorporating interference. Moreover, \textsc{beam-align} outperforms SINR-dynamic and SINR-$1$ in dense and sparse scenarios, respectively. 
Finally, we investigated the performance of OPT-SNR, \textsc{beam-align} and SINR-dynamic under scenarios with blockage, rain, and clustered users. We observed that the SINR-dynamic scheme is inferior to OPT-SNR and \textsc{beam-align}, and that \textsc{beam-align} is robust to imperfections in the operation environment.


\bibliographystyle{IEEEtran}

\end{document}